\documentclass[reprint,superscriptaddress,amsmath,amssymb,aps,twocolumn,pra]{revtex4-2}

\usepackage{color, amssymb, graphicx, amsfonts,epstopdf,float}
\usepackage{multirow,amssymb,amsbsy,amsmath,epstopdf}
\usepackage{hyperref}
\usepackage{verbatim}
\usepackage{bm,braket,comment}
\usepackage{bbold}
\usepackage{soul}
\usepackage{subfigure}

\begin{document}
	
\title{Loss and distinguishability effects in heralded entangled state generation with Gaussian resources}
	
\author{Yun-Long Cao}
\author{Xiao-Ye Xu}
\email{xuxiaoye@ustc.edu.cn}
\affiliation{Laboratory of Quantum Information, University of Science and Technology of China, Hefei 230026, China}
\affiliation{Anhui Province Key Laboratory of Quantum Network, University of Science and Technology of China, Hefei 230026, China}
\affiliation{CAS Center for Excellence in Quantum Information and Quantum Physics, University of Science and Technology of China, Hefei 230026, China}
\affiliation{Hefei National Laboratory, University of Science and Technology of China, Hefei 230088, China}
\author{Wiwittawin Sukmas}
\affiliation{Extreme Conditions Physics Research Laboratory and Center of Excellence in Physics of Energy Materials (CE:PEM), Department of Physics, Faculty of Science, Chulalongkorn University, Bangkok, 10330, Thailand}
\author{Chuan-Feng Li}
\email{cfli@ustc.edu.cn}
\author{Guang-Can Guo}
\affiliation{Laboratory of Quantum Information, University of Science and Technology of China, Hefei 230026, China}
\affiliation{Anhui Province Key Laboratory of Quantum Network, University of Science and Technology of China, Hefei 230026, China}
\affiliation{CAS Center for Excellence in Quantum Information and Quantum Physics, University of Science and Technology of China, Hefei 230026, China}
\affiliation{Hefei National Laboratory, University of Science and Technology of China, Hefei 230088, China}
	
\date{\today}
	
\begin{abstract}
The effects of optical loss and photon distinguishability on the heralded generation of entangled states based on Gaussian resources are quantitatively investigated. By incorporating mode-dependent loss and the statistical characteristics of partially distinguishable photons into a phase-space representation, an efficient numerical framework is established to optimize target-state fidelity and success probability, with a specific focus on the enhancement triggered by non-Gaussian operations such as photon addition and subtraction. The numerical optimization results indicate that a non-vacuum post-selection strategy within a dual-rail encoding framework, combined with the simultaneous tuning of squeezing parameters and the interferometer network, effectively suppresses vacuum noise, enabling the generation of high-fidelity Bell, GHZ, and W states under realistic experimental constraints. The results show that introducing non-Gaussian operations can successfully enhance the state generation performance under realistic imperfections analogous to the enhancements observed under ideal conditions. This study demonstrates that experimental imperfections primarily scale down the success probability rather than fundamentally compromising the state fidelity, providing practical design guidelines for scalable state engineering on integrated photonic platforms.		
\end{abstract}
	
\maketitle
	
\section{Introduction}

Quantum entanglement, a hallmark distinction between quantum and classical systems\,\cite{RevModPhys.81.865,Nielsen_Chuang_2010,RevModPhys.86.419}, is widely recognized as a foundational resource for a wide range of quantum applications, including quantum computation, quantum communication, and quantum metrology\,\cite{PhysRevLett.67.661,PhysRevLett.70.1895,PhysRevLett.86.5188,doi:10.1126/science.1104149}. As for the photonic quantum information, entangled states encoded on the Fock basis can be generated via conditional measurements on Gaussian states\,\cite{cao2026heraldedentangledstategeneration,doi:10.1126/science.aay4354,PhysRevA.100.052301,PhysRevA.109.023717}, providing a scalable and experimentally feasible route toward heralded quantum resource engineering for these applications.

Despite their theoretical promise, most previous theoretical proposals for the heralded generation of non-classical and entangled states rely on idealized assumptions\,\cite{PhysRevResearch.3.043031,PhysRevA.102.012604,PhysRevA.100.052301,PhysRevA.59.1658,doi:10.1126/science.1122858,PhysRevA.80.053822,PhysRevA.86.012328,doi:10.1126/science.aay4354,PhysRevA.109.023717}, such as pure single-mode Gaussian sources, lossless linear interferometers and detectors, and perfectly indistinguishable photons. In practical photonic platforms, however, optical loss, mode mismatch, and partial photon distinguishability are unavoidable, and these experimental imperfections can significantly degrade both the success probability and the fidelity of the heralded states\,\cite{PhysRevLett.132.130603,PhysRevLett.132.130604,PhysRevLett.125.110506,ZHONG2019511}. Therefore, a quantitative study of these imperfections is of paramount importance for evaluating the feasibility of Gaussian-based generation of non-classical and entangled states in near-term experiments.

Prior studies in this context have primarily focused on the classical simulation of non-ideal Gaussian Boson Sampling (GBS)\,\cite{PhysRevLett.124.100502,oh2024classical,shi2022effect}, aiming to verify whether quantum systems maintain their quantum advantage under experimental constraints. These models simulate loss and distinguishability by introducing mode-dependent Gaussian operations, revealing that non-ideal conditions fundamentally alter the computational complexity of classical simulation and thus determine the threshold for quantum advantage.

In this work, we develop a comprehensive numerical framework to simulate the heralded generation of entangled states from multimode Gaussian sources under realistic experimental imperfections. By incorporating mode-dependent loss channels and the statistical characteristics of distinguishable photons, and leveraging the properties of Gaussian states in the phase-space representation, we ensure both the efficiency and computational feasibility of the model. We design a conditional probability structure that enables efficient and accurate evaluation and optimization of the heralded target-state fidelity and success probability. By simultaneously optimizing the squeezing parameters and the interferometer network, we demonstrate that high-fidelity many-body entangled states can be generated even in the presence of moderate loss and photon distinguishability. This framework provides practical design guidelines for integrated photonic platforms and offers a scalable tool for realistic resource estimation and optimization of Gaussian-based non-classical and entangled state engineering, thereby bridging the gap between idealized theoretical proposals and experimental implementations.

\section{Background}
In the standard framework of quantum mechanics, a composite system is defined within a Hilbert space $\mathcal{H}$ that is the tensor product of $n$ subsystem spaces, $\mathcal{H}=\otimes_{l=1}^n \mathcal{H}_l$. A general pure state within this space can be expressed as:\begin{equation}\ket{\psi}=\sum_{i_1,\ldots,i_n} c_{i_1,\ldots,i_n} \ket{i_1}\otimes\ket{i_2}\otimes\cdots\otimes\ket{i_n}.\end{equation}The state is classified as separable if it can be factorized into a product of individual subsystem states, $\ket{\psi}=\otimes_{l=1}^n \ket{\psi_l}$; otherwise, the state is entangled, exhibiting non-classical correlations that inherently prevent it from being prepared from separable states via local operations and classical communication (LOCC)\,\cite{RevModPhys.81.865}. In photonic implementations, target resources such as Bell, Greenberger–Horne–Zeilinger (GHZ), and W states\,\cite{PhysRevLett.23.880,Greenberger1989,PhysRevA.62.062314} represent different classes of multipartite entanglement encoded on the Fock basis. While theoretical generation schemes typically rely on idealized pure-state transformations to overcome this LOCC constraint, realistic experimental implementations inevitably introduce non-ideal factors such as photon loss and partial distinguishability. The presence of such noise transforms the ideal pure states into mixed states, degrading the entanglement of the system\,\cite{Aolita_2015}. Consequently, the preparation of multipartite entanglement faces a critical challenge: it must transcend mere probabilistic generation to establish practical protocols capable of preserving non-separability under realistic experimental imperfections.

In continuous-variable (CV) quantum systems, multi-mode Gaussian states are a category of great concern because the information is encoded in the quadratures of bosonic modes\,\cite{RevModPhys.84.621}. An N-mode bosonic system is described by the quadrature operators $\hat{\bm{x}}=(\hat{q}_1,\hat{p}_1,\dots,\hat{q}_N,\hat{p}_N)^T$, which are defined via the creation and annihilation operators. A key advantage of the Gaussian formalism is that such states are fully characterized by their first and second moments, namely the displacement vector \(\bar{\bm{x}}=\langle \hat{\bm{x}} \rangle\) and the covariance matrix \(\bm{V}\) with elements \(V_{ij}=\tfrac{1}{2}\langle \hat{x}_i\hat{x}_j+\hat{x}_j\hat{x}_i \rangle-\langle\hat{x}_i\rangle \langle\hat{x}_j\rangle \). In phase space, this leads to a quasi-probability representation via the Wigner function, which for Gaussian states takes the form of a multivariate Gaussian distribution: 
\begin{equation}
	W_G(\bm{x}) =
	\frac{\exp\!\left[-\tfrac{1}{2}(\bm{x}-\bm{\bar{x}})^T \bm{V}^{-1}(\bm{x}-\bm{\bar{x}})\right]}
	{(2\pi)^N \sqrt{\det\bm{V}}} .
\end{equation}

Gaussian operations, which map Gaussian states to other Gaussian states, are highly compatible with this phase-space representation. Unitary transformations generated by quadratic Hamiltonians correspond to symplectic transformations $\bm{S}$ and displacements $\bm{d}$ under which the moments evolve as $\bm{\bar{x}}\rightarrow \bm{S}\bm{\bar{x}}+\bm{d}$ and $V\rightarrow \bm{S}\bm{V}\bm{S}^T$. Furthermore, non-unitary operations such as partial tracing, mathematically equivalent to removing specific degrees of freedom in phase space, preserve the Gaussian nature of the remaining subsystems.\,\cite{PRXQuantum.2.030204}. The property under both unitary and non-unitary Gaussian operations provides a robust framework for modeling the propagation of quantum states through complex optical networks.

Performing local photon-number measurements on a multi-mode Gaussian state yields a probability distribution on the Fock basis. For a Gaussian state with a covariance matrix \(\bm{V}\) and zero displacement, the probability of a specific measurement outcome $\bm{n}$ is
\begin{equation}
	P(n) = \frac{\mathrm{Haf}(A_S)}{\bm{n}!\sqrt{|V_Q|}},
\end{equation}
where $\bm{V}_Q = \bm{V} + \bm{I}/2$ and $\bm{A}_S$ is a submatrix of $\bm{A} = \bm{X}(\bm{I} - \bm{V}_Q^{-1})$ chosen according to the detection pattern. The Hafnian is a matrix function summing over all perfect matching permutations. For states with a non-zero displacement $\bar{\bm{x}}$, the probability $P(\bm{n})$ incorporates additional coherent contributions via a displacement-dependent expansion. Due to the \#P-class computational complexity of the Hafnian, this framework serves as the theoretical foundation for demonstrating quantum advantage.

\begin{figure}[h]
	\centering
	\includegraphics[width=0.35\textwidth]{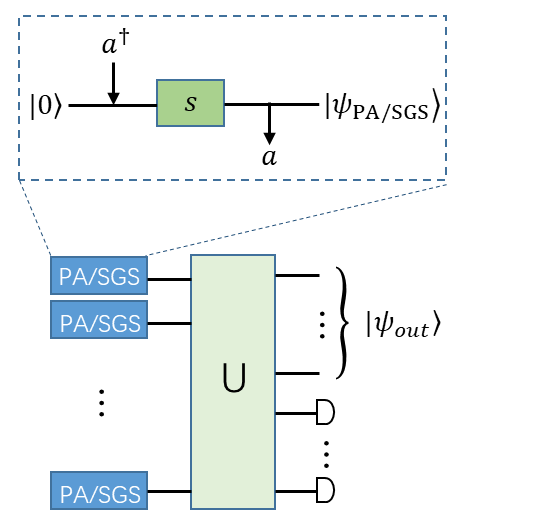}
	\caption{Schematic of ideal heralded entanglement generation model with photon addition/subtraction Gaussian source. The photon addition and subtraction operations are realized via Gaussian operations and conditional measurements, which can be arbitrarily configured as required. The single-mode squeezers S and the linear interferometer U are serving as optimization parameters.}
	\label{fig:Model_ideal}
\end{figure}

In the context of state engineering, the measurement process enables a heralding strategy\,\cite{Forbes_2025}. By conditioning on specific outcomes in a subset of ancillary modes, a target entangled state can be probabilistically heralded in the remaining modes. For an initial pure state $\ket{\Psi}$, the successful detection of a herald signal $\bm{h}$ with probability $p$ yields the desired resource $\ket{\phi}$ via the decomposition:
\begin{equation}
	\ket{\Psi} = \sqrt{p}\ket{\phi}\ket{\bm{h}} + \sqrt{1-p}\ket{R},
\end{equation}
where $\ket{R}$ represents the discarded component\,\cite{PhysRevResearch.3.043031}. The quality of the generation is determined by the success probability and the fidelity with target state. 

\begin{figure}[h]
	\centering
	\includegraphics[width=0.35\textwidth]{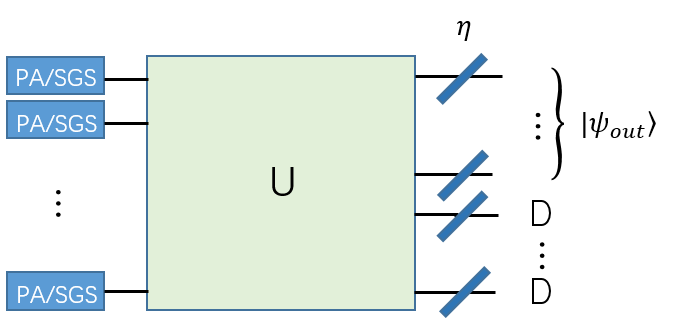}
	\caption{Schematic of heralded entanglement generation model with loss. The transmissivity $\eta$ of beam splitters is used to model the total loss. The single-mode squeezers S and the linear interferometer U are serving as optimization parameters.}
	\label{fig:Model_loss}
\end{figure}

In practical quantum optical experiments, non-ideal factors are inherent at every stage, including the preparation of Gaussian states, interferometric transformations, and measurement processes. The detrimental effects of optical loss, mode mismatch and finite detection efficiency can be categorized into two primary mechanisms: photon loss and distinguishability. Photon loss couples the ideal Gaussian state with vacuum noise, while distinguishability disrupts the quantum interference between photons\,\cite{PhysRevA.48.4598,PhysRevLett.59.2044}. Both mechanisms fundamentally alter the statistical characteristics of heralded measurements and the density matrix of the output states, leading to a significant degradation in both the success probability and the fidelity of the target states. To rigorously evaluate the impact of these imperfections on the quality of state generation, and to minimize such degradation while maintaining computational feasibility, it is essential to re-engineer the optimization models within the framework of Gaussian operations and photon-number measurements.

\begin{figure}[h]
	\centering
	\includegraphics[width=0.35\textwidth]{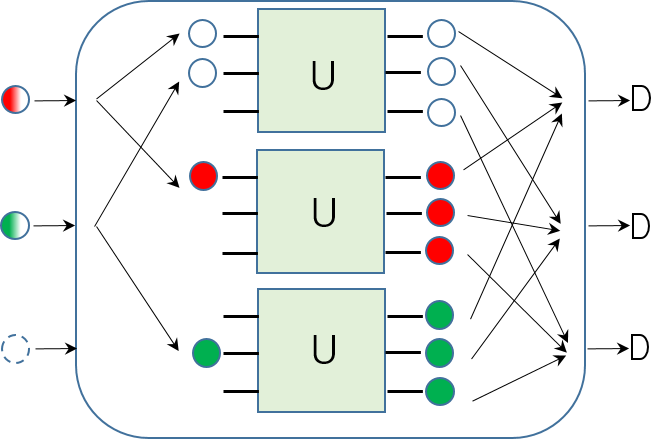}
	\caption{Schematic of heralded entanglement generation model with partial distinguishability. Partially distinguishable photon states are inputted into the first two modes of a three-mode linear interferometer. The states are decomposed into indistinguishable components and distinguishable components that propagate independently. Both pass through the interferometer to the detectors. The solid rectangle U represents the interferometer for indistinguishable photons. The dashed rectangles represent the virtual interferometer for distinguishable photons.}
	\label{fig:Model_dist}
\end{figure}

\section{Method}
The heralded generation of entangled states begins with the construction of a multi-mode Gaussian resource. The core Gaussian source is composed of multiple single-mode squeezed states. To enhance the non-Gaussianity of the source, we integrate heralded photon addition and photon subtraction modules as illustrated in Fig.~\ref{fig:Model_ideal}. Specifically, photon addition (PA) is modeled by a two-mode squeezing operation between a system mode and an ancillary mode, followed by a photon-number-resolving (PNR) detection on the ancillary mode. Photon subtraction (PS) is modeled by a beam-splitter transformation that couples a system mode with an ancillary vacuum mode, followed by a PNR detection on the ancilla\,\cite{PhysRevA.96.043861}. In our configuration, PA is applied prior to the primary squeezing stage to exploit potential stimulated effects, while PS is performed post-squeezing to mitigate vacuum-induced noise. This configuration maximizes the impact of non-Gaussian operations on the state's statistical structure. Throughout this work, photon addition and subtraction are regarded as available non-Gaussian resources, and the optimization focuses on quantifying the improvement they provide to heralded state engineering.

\begin{figure}[h]
	\centering
	\includegraphics[width=80mm]{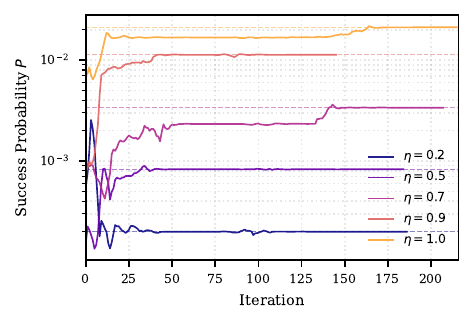}
	\\[4pt]
	\includegraphics[width=80mm]{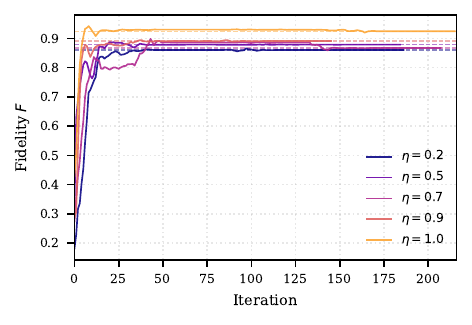}
	\caption{Probability and fidelity of the Bell state generated with different efficiency as a function of optimal steps. The weights \(w_1\), \(w_2\) and \(\epsilon\) are set to \(10\), \(1\) and \(10^{-4}\).}
	\label{fig:trace_loss_Bell}
\end{figure}
\begin{figure}[h]
	\centering
	\includegraphics[width=80mm]{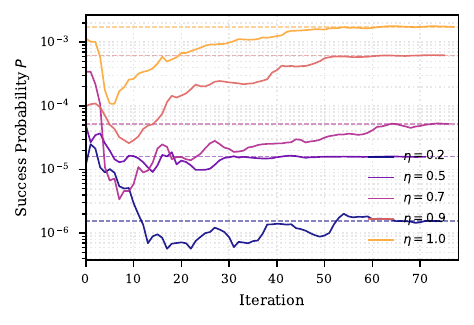}
	\\[4pt]
	\includegraphics[width=80mm]{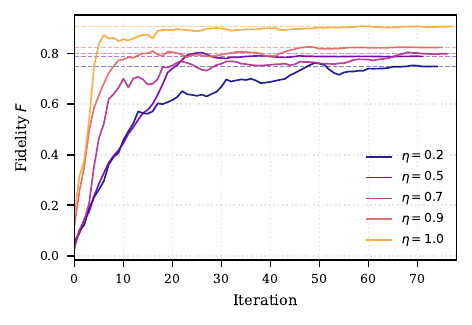}
	\caption{Probability and fidelity of the GHZ state generated with different efficiency as a function of optimal steps. The weights \(w_1\), \(w_2\) and \(\epsilon\) are set to \(10\), \(1\) and \(10^{-4}\).}
	\label{fig:trace_loss_GHZ}
\end{figure}
\begin{figure}[h]
	\centering
	\includegraphics[width=80mm]{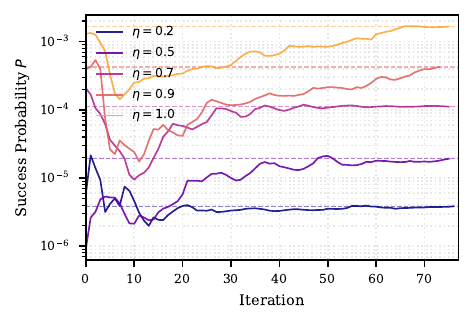}
	\\[4pt]
	\includegraphics[width=80mm]{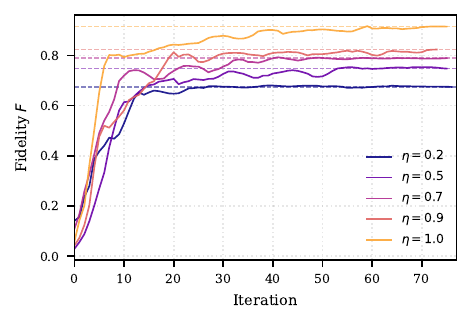}
	\caption{Probability and fidelity of the W state generated with different efficiency as a function of optimal steps. The weights \(w_1\), \(w_2\) and \(\epsilon\) are set to \(10\), \(1\) and \(10^{-4}\).}
	\label{fig:trace_loss_W}
\end{figure}
\begin{figure}[h]
	\centering
	\includegraphics[width=80mm]{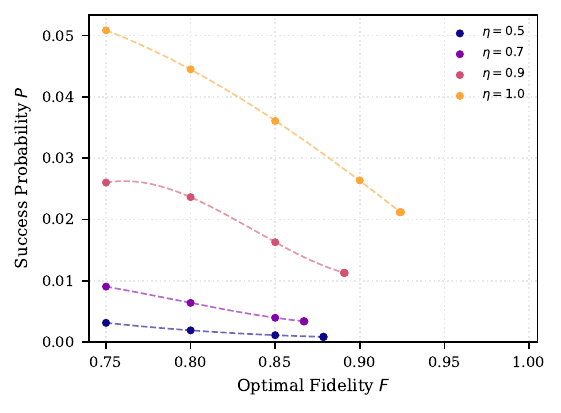}
	\caption{Optimized Pareto frontiers of the Bell state under varying degrees of photon efficiency $\eta$. The contours delineate the success probability $P$ as a function of the optimal fidelity $F$. Each dashed line represents the global trade-off boundaryfor a specific efficiency.}
	\label{fig:pareto_loss_Bell}
\end{figure}

Following the source preparation, the modes are directed into a linear interferometer, modeled as a universal unitary transformation $U$. Experimentally, this $N$-mode interferometer can be decomposed into a network of two-mode transformations, such as beam splitters and phase shifters, which serve as the primary degrees of freedom for the optimization of entanglement structures\,\cite{Clements:16}. The final stage in the heralded model is the conditional measurement on the ancillary modes. When a specific set of photon numbers is detected via PNR detectors, the remaining output modes are projected into a conditional state. If the parameters are correctly optimized, this heralded projection maps the multi-mode Gaussian-based resource onto the target entangled state, e.g., Bell, GHZ, or W states, in the Fock basis.

\begin{figure}[h]
	\centering
	\includegraphics[width=80mm]{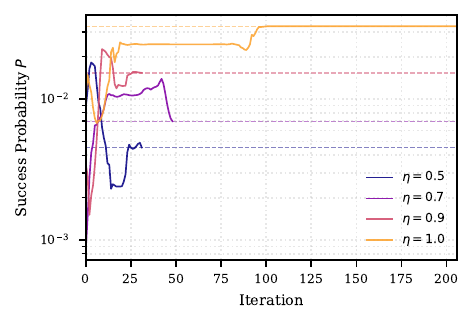}
	\\[4pt]
	\includegraphics[width=80mm]{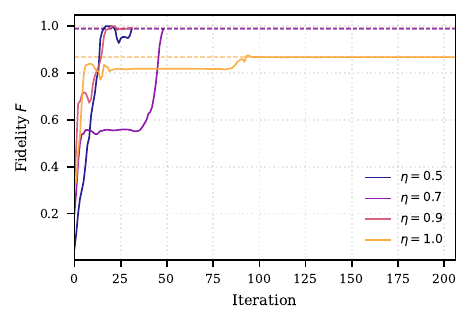}
	\caption{Probability and fidelity of the Bell state generated with different indistinguishable rates \(\eta\) as a function of optimal steps. The weights \(w_1\), \(w_2\) and \(\epsilon\) are set to \(5\), \(1\) and \(10^{-4}\).}
	\label{fig:trace_dist_Bell}
\end{figure}
\begin{figure}[h]
	\centering
	\includegraphics[width=80mm]{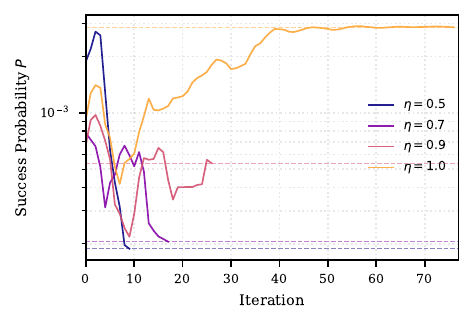}
	\\[4pt]
	\includegraphics[width=80mm]{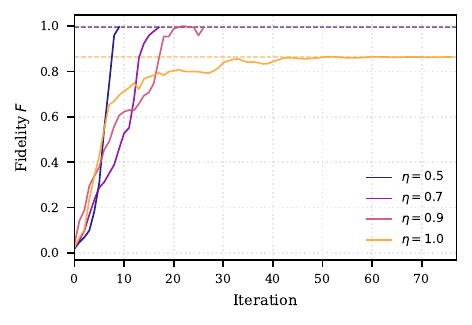}
	\caption{Probability and fidelity of the GHZ state generated with different indistinguishable rates \(\eta\) as a function of optimal steps. The weights \(w_1\), \(w_2\) and \(\epsilon\) are set to \(5\), \(1\) and \(10^{-4}\).}
	\label{fig:trace_dist_GHZ}
\end{figure}
\begin{figure}[h]
	\centering
	\includegraphics[width=80mm]{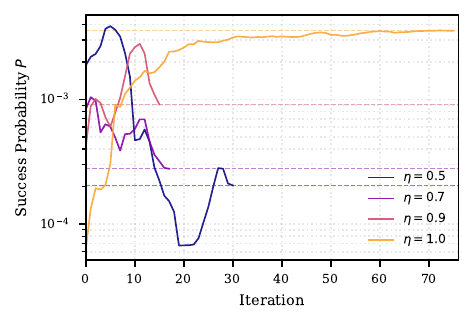}
	\\[4pt]
	\includegraphics[width=80mm]{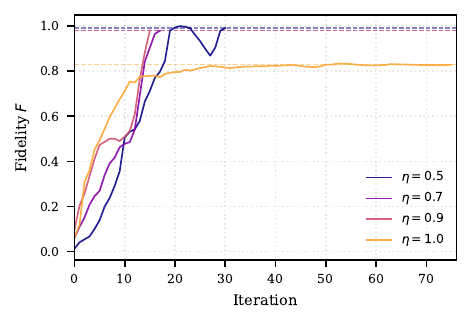}
	\caption{Probability and fidelity of the W state generated with different indistinguishable rates \(\eta\) as a function of optimal steps. The weights \(w_1\), \(w_2\) and \(\epsilon\) are set to \(5\), \(1\) and \(10^{-4}\).}
	\label{fig:trace_dist_W}
\end{figure}
\begin{figure}[h]
	\centering
	\includegraphics[width=80mm]{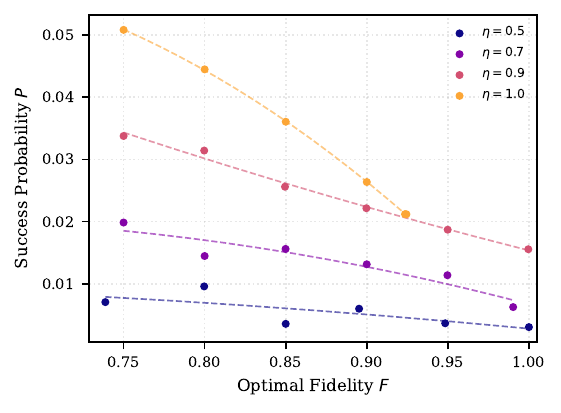}
	\caption{Optimized Pareto frontiers of the Bell state under varying degrees of photon indistinguishable rates $\eta$. The contours delineate the success probability $P$ as a function of the optimal fidelity $F$. Each dashed line represents the global trade-off boundary for a specific indistinguishable rate.}
	\label{fig:pareto_dist_Bell}
\end{figure}

A significant challenge in heralded Gaussian models is the vacuum component in the output modes, which originates from the intrinsic vacuum dominance of the squeezed resources. As noted in our previous work \cite{cao2026heraldedentangledstategeneration}, this vacuum contribution cannot be completely eliminated by a finite number of heralding or photon addition/subtraction operations, thereby limiting the maximum achievable fidelity. To mitigate this, we implement a non-vacuum qubit post-selection strategy within the dual-rail encoding framework. Each logical qubit is encoded across two physical modes, and we exclude all measurement events where any logical qubit collapses into a vacuum state. Experimentally, this can be efficiently realized using threshold detectors to ensure each qubit contains at least one photon. This strategy effectively filters out the dominant vacuum noise, projecting the state into a subspace that significantly enhances the fidelity of the generated entanglement\,\cite{PhysRevApplied.17.034071}.

To transition from an ideal theoretical model to a realistic experimental assessment, we introduce modular components to simulate photon loss and partial distinguishability. Experimental loss including propagation attenuation, coupling inefficiencies, and finite detector quantum efficiency are modeled as linear loss channels\,\cite{PhysRevLett.124.100502,oh2024classical}. We simulate this by coupling each system mode, both target and heralding modes, with an auxiliary vacuum mode via a beam splitter with transmissivity $\eta$ as illustrated in Fig.~\ref{fig:Model_loss}. The closure of Gaussian states under such Gaussian operations ensures that the reduced state in the system modes remains Gaussian, although it evolves from a pure to a mixed state.

In realistic implementations, mode mismatches in spatial, temporal, and spectral degrees of freedom cause ideal single-mode states to expand into a multimode structure, leading to a reduction in quantum interference visibility. This effect can be effectively characterized through photon distinguishability, allowing the system to be treated by separately accounting for the indistinguishable and distinguishable photon components\,\cite{shi2022effect}. The present model adopts an effective two-component description, where each photon is decomposed into a fully indistinguishable component and a fully distinguishable component as illustrated in Fig.~\ref{fig:Model_dist}. The relative weights of these two components are determined by the indistinguishable rate. The indistinguishable component follows Bose-Einstein statistics, where probabilities $P_{\text{GBS}}(\mathbf{n}_i)$ are calculated using the Hafnian-based formalism for Gaussian Boson Sampling. The distinguishable component follows classical statistics, where photons propagate independently. The probability $P_{\text{class}}(\mathbf{n}_d)$ is determined by the photon statistics of the source and the classical transmission matrix of the linear network\,\cite{10.1145/1993636.1993682,PhysRevA.91.022316}. Although a realistic mode mismatch is generally described by a continuous overlap matrix between internal photon modes, the present approximation captures the transition between perfect quantum interference and completely classical behavior while avoiding the additional complexity associated with modeling arbitrary overlap matrices. This effective description is intended to capture the dominant influence of partial distinguishability on the heralding performance, rather than to reproduce the full multimode structure of a specific experimental implementation. Since practical PNR detectors cannot distinguish these two components, the total probability $P(\mathbf{n})$ for a detected outcome $\mathbf{n}$ is obtained by summing the joint probabilities over all possible combinations:

\begin{equation}
	P(\mathbf{n}) = \sum_{\mathbf{n} = \mathbf{n}_i + \mathbf{n}_d} P_{\text{GBS}}(\mathbf{n}_i) \cdot P_{\text{classic}}(\mathbf{n}_d).
\end{equation}

The quality of the heralded entanglement is quantified by the fidelity $F$ and the success probability $P$. The success probabilities throughout this work characterize the conditional state generation performance, the availability of the required non-Gaussian resource. Consequently, the optimization isolates the influence of non-Gaussian operations on the heralded state generation process, while the preparation efficiency of the non-Gaussian resource can be considered separately for a specific experimental implementation. Under non-ideal conditions, the output is a mixed state $\rho_{out}$, and the fidelity is defined as the overlap with the ideal target vector: $F = \langle \psi_{\text{target}} | \rho_{\text{out}} | \psi_{\text{target}} \rangle$. The optimization parameters $\bm{\xi}$ consist of the squeezing parameters $r_i$ for  single-mode squeezed source and the linear interferometer parameters $(\theta, \phi)$. The cost function is defined as:
\begin{equation}
	f(\bm{\xi}) = -w_1\ln(F) - w_2\ln(P) + \epsilon \sum ||\xi_i||^2,
\end{equation}
where $w_1, w_2$ are weights balancing efficiency and quality. Numerical computations are performed using The Walrus package~\cite{Gupt2019}. 
The optimization is carried out using the L-BFGS algorithm implemented. 
Since the objective function is generally non-convex due to the nonlinear dependence of the Hafnian probabilities on the Gaussian parameters, multiple optimization runs are performed from randomly generated initial parameters, and the solution with the lowest cost is selected to mitigate convergence to local optima.

The computational cost is dominated by repeated evaluations of Hafnians associated with conditional photon-number probabilities. Consequently, the optimization complexity increases rapidly with the number of optical modes and detected photons. 
Nevertheless, by exploiting the Gaussian phase-space formalism, the computational task is reduced to repeated Hafnian evaluations of reduced Gaussian states, avoiding explicit Fock-basis expansions and truncations throughout the optimization.

\section{Results}
\subsection{Loss}
The key to realizing heralded state generation is the rapid and stable identification of system parameters that yield both high fidelity and high probability—a requirement that remains critical in loss models. We optimized the squeezing parameters and the linear interferometer of a heralded entangled state generation model subject to linear loss rates. Specifically, we recorded the loss function, fidelity, and success probability as functions of optimization steps during the generation of dual-rail encoded Bell, GHZ, and W states under varying loss rates, comparing these with the optimal results of an ideal entanglement generation model. The Bell, GHZ and W state generation models contain 6, 10 and 10 optical modes, respectively, including both target and heralding modes.

\begin{figure}[h]
	\centering
	\includegraphics[width=80mm]{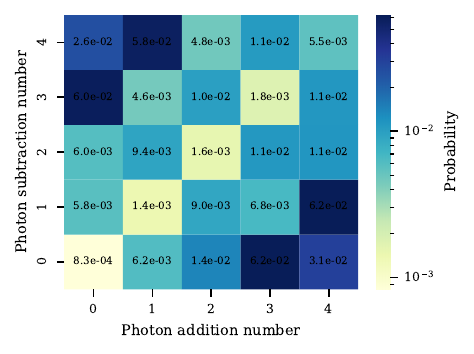}
	\\[4pt]
	\includegraphics[width=80mm]{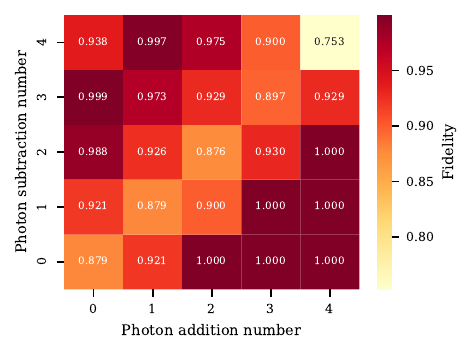}
	\caption{Probability and fidelity of the Bell state generated as a function of photon addition and subtraction with the efficiency \(\eta=0.5\).}
	\label{fig:heatmap_loss_Bell}
\end{figure}

\begin{figure}[h]
	\centering
	\includegraphics[width=80mm]{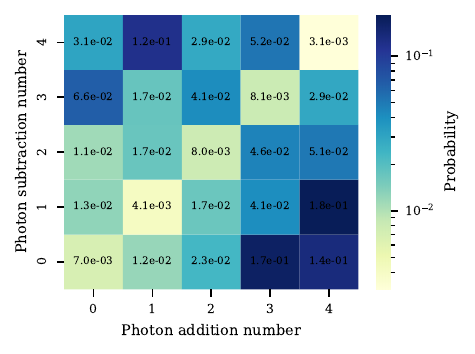}
	\\[4pt]
	\includegraphics[width=80mm]{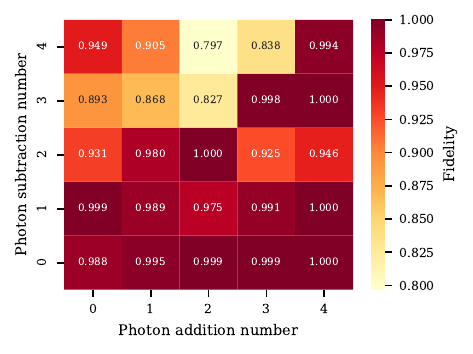}
	\caption{Probability and fidelity of the Bell state generated as a function of photon addition and subtraction with the indistinguishable rates \(\eta=0.7\).}
	\label{fig:heatmap_dist_Bell}
\end{figure}

As shown in Fig.~\ref{fig:trace_loss_Bell}, \ref{fig:trace_loss_GHZ} and \ref{fig:trace_loss_W}, the success probability of our model decays significantly as the loss rate increases, while the fidelity experiences only a moderate decrease, remaining at a relatively high level. The optimized trade-off between fidelity and success probability under various loss rates is comprehensively mapped out by the Pareto frontiers in Fig.~\ref{fig:pareto_loss_Bell}. This occurs because the loss channel mixes the ideal Gaussian state with vacuum noise, increasing the proportion of the vacuum state component. Although the energy dissipation from loss and the exclusion of vacuum components during post-selection inevitably reduce the state generation efficiency, our model mitigates the primary impact of the vacuum state on fidelity through post-selection.

\subsection{Distinguishability}
Our model separates Gaussian states according to their distinguishability using virtual beam splitters, where changes in distinguishability are simulated by adjusting the beam splitter transmissivity. We optimized the squeezing parameters and the linear interferometer for the heralded entangled state generation model under distinguishability constraints. The loss function, fidelity, and success probability are recorded as functions of optimization steps for generating dual-rail encoded Bell, GHZ, and W states under different distinguishability and compared with the ideal model. The Bell, GHZ and W state generation models contain 6, 10 and 10 optical modes, respectively, including both target and heralding modes.

As illustrated in Fig.~\ref{fig:trace_dist_Bell}, \ref{fig:trace_dist_GHZ} and \ref{fig:trace_dist_W}, while the probability decays noticeably with increasing distinguishability, the fidelity can actually exceed the upper bound found in the ideal Gaussian state model. The optimized trade-off between fidelity and success probability under various distinguishable rates is comprehensively mapped out by the Pareto frontiers in Fig.~\ref{fig:pareto_dist_Bell}. Optimization results indicate that in the presence of partial distinguishability, our model tends toward low-gain squeezing. This enhances the fidelity boost provided by post-selection and suppresses the negative impact of multi-photon terms, albeit at the cost of lower generation efficiency. Distinguishable photons follow classical statistical distributions, altering the statistical characteristics of the vacuum and multi-photon terms in the output state. Consequently, the heralding process affects distinguishable and indistinguishable photon states differently, allowing the target states (defined as indistinguishable) to be more effectively filtered into the conditional state.

\subsection{Enhanced by Non-Gaussian Operations}
The enhancement of non-Gaussianity in Gaussian sources through photon addition or subtraction is known to improve both the fidelity and the success probability of state generation under ideal conditions. We incorporate these non-Gaussian operations into our heralded entanglement generation models subject to linear loss and partial distinguishability, which allows for a comprehensive comparison of the optimized performance. Since photon addition and subtraction are regarded as representative non-Gaussian resources in this work, separating the resource preparation efficiency from the conditional heralding probability allows the intrinsic state-engineering capability of different non-Gaussian operations to be compared on an equal footing.

The optimization results for the lossy model, as illustrated in Fig.~\ref{fig:heatmap_loss_Bell}, indicate that non-Gaussian operations substantially improve the state generation performance in terms of both fidelity and success probability compared with the Gaussian-only baseline. Both photon addition and subtraction contribute constructively to enhancing the heralded state generation. In particular, a distinct numerical phenomenon observed along the diagonal of the heatmap, where the optimization efficacy asymptotically degenerates to the baseline Gaussian model. This behavior can be rigorously explained by the operator symmetry erasure in the low squeezing regime. To illuminate this mechanism, we examine a single-photon subtraction acting on a single-mode squeezed vacuum state $\hat{S}(r)\ket{0}$, which yields
\begin{equation}
	\hat{a}\hat{S}(r)\ket{0} = -\sinh r\hat{S}(r)\ket{1}.
\end{equation}
This single-photon subtraction breaks the even parity of the initial state, transforming it into a single-term, odd-parity non-Gaussian state, similarly to the case of a single-photon addition\,\cite{Kim_2008}. However, the combined and balanced action of both photon addition and subtraction operations results in a coherent two-term superposition:
\begin{equation}
	\hat{a}\hat{S}(r)\hat{a}^\dagger\ket{0} = \cosh r\hat{S}(r)\ket{0}-\sqrt{2}\sinh r\hat{S}(r)\ket{2}.
\end{equation}
In the low squeezing regime, the second term carrying the non-Gaussian coherence is heavily suppressed, meaning that the Gaussian component rapidly becomes predominant. Consequently, the balanced non-Gaussian operations effectively reduce to a unity operation, leading to the performance erosion along the heatmap diagonal.

In the distinguishability model as shown in Fig.~\ref{fig:heatmap_dist_Bell}, the optimization without non-Gaussian enhancement tends toward near-unity fidelity due to the influence of the incoherent photon distribution. In this regime, the primary role of photon addition or subtraction shifts toward enhancing the state generation probability relative to Gaussian-only implementations. The results indicate that while incorporating non-Gaussian operations effectively increases the generation probability, photon subtraction leads to a degradation in fidelity.

The presence of distinguishability drives the optimization toward a low squeezing regime. Under such conditions, the squeezed state following a photon-subtraction operation tends to enhance the single-photon component, which facilitates fidelity improvement for targeting dual-rail encoded entangled states. In the present framework, photon subtraction is treated as a deterministic operation in the state evolution within the simulation framework, with the success probability of the subtraction channel factored out from the optimization objective. As a result, the usual trade-off between fidelity and success probability is partially suppressed, and the optimization is biased toward increasingly low squeezing parameters. However, to ensure the experimental feasibility of the optimized results, the squeezing parameters are constrained by a lower bound. Consequently, the optimization may converge to a state with lower fidelity. 

In contrast, photon addition does not induce a comparable drift toward vanishing squeezing regimes, since its success probability is independent of the squeezing parameters and therefore does not bias the optimization toward lower squeezing parameters. As a result, it provides a more stable route for enhancing non-Gaussian features under experimental constraints, leading to improved robustness in both fidelity and success probability within the distinguishability model.

\section{Discussion}
By employing on-off post-selection on the qubit modes, our model effectively mitigates the impact of the vacuum component inherent in Gaussian states. This approach provides a similar suppression effect on the vacuum noise introduced by external factors such as linear loss or partial photon distinguishability. Consequently, the degradation of state fidelity under these non-ideal conditions remains significantly less pronounced than the corresponding noise intensity. The dual-rail encoded entangled states employed in our model are highly compatible with local post-selection operations. In practice, this post-selection process can be implemented through various established techniques, such as quantum nondemolition measurements or simple threshold detectors, depending on the specific requirements of the target state. Such versatility ensures the experimental feasibility of our scheme. Our results even suggest that the classical incoherent statistical characteristics of distinguishable photons can be leveraged to surpass the fidelity limits inherently associated with ideal Gaussian sources, albeit at the cost of an inevitable reduction in generation efficiency. The optimization data demonstrate that observable multi-photon entangled states can be generated even at very low transmission efficiencies. Furthermore, in regimes with high proportions of distinguishable photons, the system can still produce indistinguishable entangled states with near-unity fidelity at the output ports.

These findings underscore the robustness of the heralded entanglement generation scheme, where the adverse effects of loss and distinguishability are primarily manifested as a reduction in success probability rather than a fundamental degradation of state fidelity. The device can maintain high-fidelity performance by appropriately tuning the system parameters without requiring additional non-Gaussian resources or complex hardware compensation. For the rapidly advancing field of photonic integrated circuits, this implies that high-fidelity entanglement generation is achievable within existing technological scales by evaluating source quality and optimizing experimental parameters accordingly. This reinforces the experimental viability of integrated quantum photonics, broadens its application scope, and provides a practical roadmap for future implementation. While this study focuses on dual-rail encoded discrete-variable entangled states, exploring whether similar methodologies can guarantee high-quality results for other target states or non-Gaussian gate operations under non-ideal conditions remains a valuable and challenging direction for future research.

\begin{acknowledgments}
This work was supported by Quantum Science and Technology-National Science and Technology Major Project (No.\,2021ZD0301200), National Natural Science Foundation of China (Nos.\,12474494, 12204468).	
\end{acknowledgments}
	
\bibliographystyle{apsrev4-2}
\bibliography{refs}
	
\end{document}